\documentclass[preprint,aps,prd,nofootinbib,preprintnumbers]{revtex4-1}
\pdfoutput=1

\usepackage{amsmath,amssymb,amsfonts,dcolumn,color,graphicx,graphics,latexsym,placeins,epsfig}
\usepackage{bm}
\usepackage{slashed}
\usepackage{natbib}
\usepackage{url}
\usepackage{psfrag}
\usepackage{tabularx}
\usepackage{hyperref}
\usepackage{cleveref}
\hypersetup{colorlinks=true}
\usepackage{comment}
\usepackage[normalem]{ulem} % put in preamble

\newcommand{\be}{\begin{equation}}
\newcommand{\ee}{\end{equation}}
\newcommand{\ba}{\begin{eqnarray}}
\newcommand{\ea}{\end{eqnarray}}

\begin{document}
  
\preprint{IPPP/26/07}
  
\title{Photon-dark photon oscillation in M87 and Crab Nebula environments}
\author{Tanmay Kumar Poddar$^{1}$\footnote{tanmay.k.poddar@durham.ac.uk}}
\author{Sourov Roy$^{2}$\footnote{tpsr@iacs.res.in}}
\author{Pratick Sarkar$^{2}$\footnote{spsps2523@iacs.res.in}}
\affiliation{$^{1}$ Institute for Particle Physics Phenomenology (IPPP), Department of Physics, Durham University, Durham DH1 3LE, United Kingdom}
\affiliation{$^{2}$ School of Physical Sciences, Indian Association for the Cultivation of Science, 2A \& 2B Raja S.C Mullick Road, Kolkata-700032, India}

\begin{abstract}

Compact astrophysical systems such as neutron stars and black holes provide powerful laboratories for testing feebly coupled dark photons (DPs). We investigate light DPs kinetically mixed with the visible photon that need not be the dark matter, focusing on resonant photon-DP oscillations in magnetized, modeled plasma environments. We show that realistic non-monotonic plasma density profiles generically enhance resonant conversion relative to monotonic models, leading to substantially stronger constraints on the photon-DP kinetic mixing parameter ($\epsilon$).
Using spectral data from the supermassive black hole (SMBH) M87*, extending to the LOFAR band, we derive a bound $\epsilon \simeq 7\times10^{-6}$ at the DP mass $m_{A'} \simeq 5\times10^{-7}\,\mathrm{eV}$ for oscillation distance $3r_{\rm ph}$, where $r_{\rm ph}$ denotes the photon sphere radius. From the Crab pulsar-wind Nebula, we obtain an even stronger constraint, $\epsilon \simeq 8\times10^{-7}$ at $m_{A'} \simeq 4\times10^{-9}\,\mathrm{eV}$ for oscillation baselines of order $10^{3}\,\mathrm{km}$, surpassing existing astrophysical limits in realistic plasma backgrounds. While laboratory and cosmological bounds remain slightly stronger at comparable masses, observation of compact objects with larger surface magnetic fields and measurements of photon spectra at lower frequencies would enhance the limits on the photon-DP coupling by orders of magnitude.

\end{abstract}

\maketitle

\section{Introduction}
Compact objects such as black holes (BHs) and neutron stars (NSs) provide unique cosmic laboratories for exploring physics beyond the Standard Model (SM) through their observed electromagnetic (EM) emission spectra \cite{Baryakhtar:2022hbu,Bramante:2023djs}. Compared to conventional terrestrial experiments, these astrophysical systems offer several advantages in probing ultralight beyond-SM degrees of freedom, such as dark photons (DPs) that could constitute a fraction of the dark matter (DM) in the Universe. The extreme conditions in and around compact stars characterized by intense magnetic fields, high particle densities, and large spatial extensions significantly enhance the interaction probability of such light particles with the surrounding plasma and radiation. Moreover, their long-lived and continuously monitored emission allows for the accumulation of small effects over extended timescales. Consequently, compact stars act as natural, large-volume detectors that provide a powerful and complementary avenue to Earth-based direct detection experiments in the search for light dark sector particles.

In contrast to axions or scalar fields \cite{Marsh:2015xka,Carenza:2024ehj,Caputo:2024oqc,Hui:2016ltb,Poddar:2025oew}, DPs represent a spin-$1$ extension of the SM, interacting with ordinary photons through kinetic mixing \cite{Fabbrichesi:2020wbt}. The vector nature of DPs leads to distinct phenomenology. Unlike photon-axion oscillations, which require a non-zero $ \mathbf{E}\cdot\mathbf{B}$ background from the stellar electric $(\mathbf{E})$ and magnetic $(\mathbf{B})$ fields, photon-DP conversion arises through plasma-induced resonances that naturally occur in compact-star magnetospheres. Moreover, the accessible mass range and coupling structure of DPs are less model-dependent than those of axions, allowing a broader exploration of hidden-sector parameter space. Astrophysical environments such as NSs, BHs and magnetars, with their magnetic fields and dense plasmas, therefore offer robust means to probe DP couplings, providing powerful and complementary constraints to those obtained from terrestrial axion and scalar searches.

DPs can be produced through a variety of mechanisms depending on the ambient plasma conditions and the kinetic-mixing strength $(\epsilon)$ with the SM photon. In stellar and magnetized environments, their production primarily proceeds via photon-DP oscillations in regions where the effective plasma mass of the photon, $\omega_p$, approaches the DP mass, $m_{A^\prime} $, leading to resonant conversion \cite{Caputo:2020rnx}. In dense or hot plasmas, additional non-resonant production channels arise from processes such as plasmon decay \cite{An:2013yfc,Zink:2023szx}, Compton-like scattering \cite{Hochberg:2021zrf,Su:2021jvk}, and bremsstrahlung emission \cite{Tantirangsri:2023syc} involving charged particles. The corresponding emission rate depends sensitively on the local temperature, density, and magnetic field strength, which determine both the photon dispersion relation and the effective mixing angle. In compact stars, the combination of high plasma frequency and intense magnetic fields enhances these production channels, allowing even feeble kinetic mixing parameters to leave observable imprints on the emitted EM spectra or energy-loss rates. Consequently, DP production in such environments provides an astrophysically powerful probe of hidden-sector physics.

In this work, we investigate photon-DP oscillations in compact-star (BH and NS) environments within a model-agnostic framework. DPs generally arise from an additional Abelian gauge symmetry, $U(1)_D$, that kinetically mixes with the SM photon through the term $(\epsilon/2)F_{\mu\nu}{F^\prime}^{\mu\nu}$, where $F_{\mu\nu}$ $(F^\prime_{\mu\nu})$ denotes the EM (dark EM) field strength tensor \cite{Holdom:1986eq,Aldazabal:2000sa,Batell:2005wa,Abel:2008ai,Acharya:2016fge,Acharya:2017kfi,Gherghetta:2019coi}. Their mass can originate either from a St\"{u}ckelberg mechanism or from a dark-Higgs field acquiring a vacuum expectation value. Depending on the mixing strength and thermal history of the Universe, DPs may be produced through freeze-in \cite{Heeba:2019jho}, freeze-out \cite{DiazSaez:2024dzx}, or misalignment processes \cite{Zhang:2025pgb}, potentially contributing to the DM density. In compact stars, the efficiency of photon-DP conversion is governed by the interplay between the DP mass $m_{A^\prime}$ and the local plasma frequency $\omega_p(r)$. When the resonance condition $m_{A^\prime} \simeq \omega_p(r)$ is met within the stellar magnetosphere or interior, part of the EM flux can convert into DPs, giving rise to observable effects such as spectral distortions \cite{Arsenadze:2024ywr}, spin correlation \cite{Feng:2025gji}, or excess cooling signatures \cite{Lasenby:2020goo}.

The effective mass of a photon can vary spatially or temporally as it propagates through the environment of a compact star \cite{Kapusta:2006pm,Heisenberg:1936nmg}. Similarly, the DP may acquire medium-dependent corrections to its effective mass from interactions with background dark-sector particles charged under $U(1)_D$ \cite{Feng:2015hja,Berlin:2022hmt,Berlin:2023gvx}. The non-adiabatic photon-DP transition probability near a level crossing--the spatial point where the photon plasma mass equals the DP mass, leading to resonant mixing is typically evaluated using the Landau-Zener (LZ) formalism \cite{Zener:1932ws,Landau:1932vnv}, whose leading order contribution follows from the stationary-phase approximation. The LZ probability is intrinsically oscillatory, making high-precision numerical evaluation challenging. However, close to resonance points, the oscillations become slow, and the relative phase between successive resonances can be averaged out, yielding a dominant stationary-phase transition probability. Nevertheless, it has been shown in \cite{Brahma:2023zcw} that the LZ approximation breaks down when the resonant conversion occurs near a local extremum of the in-medium SM photon mass. Such situations arise naturally when the background particle density of the compact star plasma varies non-monotonically in space or time along the photon trajectory.

Existing constraints on the photon-DP kinetic mixing parameter arise from a wide range of laboratory, astrophysical, and cosmological observations \cite{Caputo:2021eaa,AxionLimits}. Laboratory limits include precision tests of Coulomb's law and bounds on the photon mass \cite{Jaeckel:2010xx,Kroff:2020zhp}, and light-shining-through-walls or helioscope experiments \cite{Povey:2010hs,OShea:2023gqn,Redondo:2008aa}. Additional constraints are derived from BH superradiance \cite{Cardoso:2018tly}, reactor neutrino data \cite{Danilov:2018bks}, and cosmological observations such as COBE/FIRAS \cite{Chluba:2024wui}, Planck+unWISE \cite{McCarthy:2024ozh}, as well as from astrophysical systems including the Sun \cite{Vinyoles:2015aba}, planets \cite{Fischbach:1994ir,Yan:2023kdg}, NSs \cite{Hong:2020bxo}, and red giants \cite{Li:2023vpv}. Some of these bounds assume that DPs constitute the DM \cite{Arias:2012az,McDermott:2019lch,Witte:2020rvb,Caputo:2020bdy}. Other related constraints and projected sensitivities to the kinetic-mixing parameter have also been discussed in \cite{Caputo:2021eaa,AxionLimits}.

In this work, we derive new limits on the photon-DP kinetic-mixing parameter by analyzing low-frequency radio observations of the galaxy Messier 87 (M87) obtained with the LOw Frequency ARray (LOFAR) telescope \cite{10.1093/mnras/stv510,Heesen_2019} in the $10-1000$ MHz range, considering different plasma density profiles around its supermassive black hole (SMBH) \cite{Braude1969TheSO,Bridle1968OBSERVATIONSOR,Roger1969SpectralFD,Viner1975263MHzRS,1969ApJ...157....1K,1990PKS...C......0W,1980MNRAS.190..903L,Scaife:2012mt,deGasperin:2012id}. We further explore the global spectral energy distribution (SED) of the Crab Nebula \cite{Buhler:2013zrp} in the 1-100 GHz frequency range, incorporating polarization data from the Planck satellite \cite{Macias-Perez:2008uaf,Meyer:2010tta,Yuan:2010wa}, to place complementary constraints. Because the background plasma potential in these environments can exhibit non-monotonic variations, the photon-DP conversion probability may receive significant corrections relative to existing estimates. Such effects can also be generalized to other two-state oscillation systems, including photon-axion and neutrino flavor oscillations in matter. Importantly, the bounds obtained in this study do not rely on the assumption that the DP constitutes the DM component.

The paper is structured as follows. In Section.~\ref{sec2}, we review the formalism of photon-DP conversion in both monotonic and non-monotonic background potentials relevant to compact star environments. Section.~\ref{sec3} presents the electron density profile around M87 and photon-DP oscillation probabilities in monotonic and non-monotonic potentials. In Section.~\ref{sec4}, we perform a similar analysis using the spectral energy distribution of the Crab Nebula to obtain conversion probability. In Section.~\ref{SPEC_Modeling}, we summarize the methodology for the spectral modeling and the data analysis. Section.~\ref{sec_limits} discusses about the limits obtained on photon-DP kinetic mixing. Finally, Section.~\ref{sec7} summarizes our findings and outlines potential implications for future studies.

Throughout this paper, we adopt natural units with $c=\hbar=1$, where $c$ denotes the speed of light in vacuum and $\hbar$ is the reduced Planck constant, unless stated otherwise.

\section{Photon-dark photon oscillation probability for monotonic and non-monotonic potentials}\label{sec2}

In this section, we briefly review the photon-DP oscillation formalism and derive the standard LZ transition probability for a monotonic background potential \cite{Landau:1932vnv,Zener:1932ws}, as well as the corresponding expression for a non-monotonic potential when the resonant conversion occurs near a local extremum of the in-medium photon mass \cite{Brahma:2023zcw}. These probability formulations will be used in the subsequent analysis to study photon-DP conversion in compact star environments.

The low energy effective Lagrangian describing the photon and DP fields coupled through kinetic mixing is given by
\begin{equation}
\mathcal{L} = -\frac{1}{4}F_{\mu\nu}F^{\mu\nu} - \frac{1}{4}F^\prime_{\mu\nu}{F^\prime}^{\mu\nu}+
 \frac{\epsilon}{2}F_{\mu\nu}{F^\prime}^{\mu\nu}
 +\frac{1}{2}m_{A^\prime}^{2}A^\prime_{\mu}{A^\prime}^{\mu}
+ j^{\mu}A_{\mu},
  \label{app1}
  \end{equation}
 where $A_{\mu}$ and $A'_{\mu}$ denote the SM photon and DP fields, respectively, and $j_{\mu}$ is the SM-EM current. The kinetic-mixing term can be eliminated by transforming $A_\mu$ to the interaction basis
\begin{equation}
  {A^{\mu}} \equiv \begin{pmatrix}A^{\mu}_{a} \\A^{\mu}_{n}
  \end{pmatrix}
\end{equation}
in which the sterile state $A^{\mu}_{n}$ does not couple directly to the SM current, while the mass matrix becomes non-diagonal, allowing for oscillations between the active $ A^{\mu}_{a}$ and sterile $A^{\mu}_{n}$ photon states. Accordingly the fields transform as $ A^{a}_{\mu} \to A_{\mu} $ and $ A^{n}_{\mu} \to A'_{\mu} - \epsilon A_{\mu} $.

In the relativistic limit, $\omega \simeq k \gg m_{\gamma}, m_{A'}$, where $\omega$ and $k$ denote the frequency and wave number of the photon field $A_{\mu}$ and the photon acquires an effective mass $m_{\gamma}$ due to its propagation through a plasma of free charges and EM fields, the evolution of the transverse modes of the photon-DP system can be described by a two-state Schr\"odinger-like equation,
\begin{equation}
i\partial_z A = H A, \qquad H = H_0 + H_1,
\label{app4}
\end{equation}
where the Hamiltonian components are given by
\begin{equation}
H_0(z) =
\begin{pmatrix}
\omega + \delta(z) & 0\\
0 & \omega + \delta_{A^\prime}
\end{pmatrix},
\qquad
H_1 =
\begin{pmatrix}
0 & \epsilon\delta_{A^\prime}\\
\epsilon\delta_{A^\prime} & 0
\end{pmatrix},
\label{app5}
\end{equation}
with $\delta(z) = -m_{\gamma}^2(z)/2\omega$ and $\delta_{A^\prime} = -m_{A^\prime}^2/2\omega$.
Since $H_0(z)$ is diagonal for all $z$, it commutes with itself at unequal spatial points, i.e. $[H_0(z),,H_0(z^\prime)] = 0$.

The interaction picture Hamiltonian is 
\begin{equation}
H_\mathrm{int}(z)= U^\dagger(z) H_1(z) U(z)=\epsilon \delta_{A^\prime} \begin{pmatrix}
0 & e^{i\varphi (z)}\\
e^{-i \varphi(z)} & 0
\end{pmatrix},
\label{app7}
\end{equation}
where 
\begin{equation}
U(z)=\exp\Big[-i\int_{z_0}^z dz^\prime H_0(z^\prime)\Big]=\begin{pmatrix}
e^{-i\int (\omega+\delta)dz'}&0\\
0 & e^{-i\int (\omega +\delta_{A^\prime})dz'}
\end{pmatrix}; \quad \varphi(z)=\int ^z_{z_0}dz^\prime \delta _{\mathrm{osc}}(z^\prime),
\label{app6}
\end{equation}
with $\delta_{\mathrm{osc}}=\delta(z) -\delta_{A^\prime}$, where $z_0$ is the point where we fix our initial condition $A(z_0)=A_{\mathrm{int}}(z_0)$. Therefore, in the Schr\"odinger picture, the evolution equation of $A_\mu$ becomes
\begin{equation}
A(z)=e^{-i\int ^z_{z_0}dz^\prime H_0(z^\prime)} e^{-i\int ^z_{z_0}dz^\prime H_{\mathrm{int}}(z^\prime)}A(z_0). 
\label{app15}
\end{equation}
Therefore, in the limit $\epsilon\ll 1$, Eq.~\ref{app15} becomes
\begin{equation}
A(z)=e^{-i\int (\omega+\delta(z))dz'}\begin{pmatrix}
1 & -i\epsilon \mathcal{C}_+\\
-i\epsilon \mathcal{C}_- e^{i\varphi (z)} & e^{i\varphi (z)}
\end{pmatrix}A(z_0) +\mathcal{O}(\epsilon^2). 
\label{app20}
\end{equation}
where
\begin{equation}
\mathcal{C}_{+(-)}=\int ^z_{z_0}dz^\prime e^{+(-)i\varphi(z^\prime)}\delta _{A^\prime}(z^\prime). 
\label{app19}
\end{equation}
Thus, if the initial state at $z_0$ is purely active i.e., $A(z_0)=(1~~0)^T$, then the active to sterile transition amplitude is 
\begin{equation}
\mathcal{A}_{a \to n} (z)=-i\epsilon \mathcal{C}_-e^{i\varphi(z)}=-i\epsilon e^{i\varphi(z)}\int ^z_{z_0}dz^\prime \delta_{A^\prime}(z^\prime)e^{-i\varphi(z^\prime)}.   
\label{app21}
\end{equation}    
As, $\delta_\mathrm{osc}$ is real, $\varphi$ is real and hence $\mathcal{C}_-=\mathcal{C}^*_+$. Therefore, the active to sterile oscillation probability transition is 
\begin{equation}
P_{a\to n}=|\mathcal{A}_{a\to n}|^2=\epsilon^2\Big|\int ^{z}_{z_0}dz^\prime \delta_{A^\prime}(z^\prime)e^{i\varphi(z^\prime)} \Big|^2 +\mathcal{O}(\epsilon^3). \label{app22}  
\end{equation}
If the initial state is sterile i.e., $A(z_0)=(0~~1)^T$, then 
\begin{equation}
\mathcal{A}_{n\to a}(z)=-i\epsilon \mathcal{C}_+, ~~~P_{n\to a}=\epsilon ^2 |\mathcal{C}_+|^2=\epsilon ^2|\mathcal{C}_-|^2 +\mathcal{O}(\epsilon^3)=P_{a\to n}. 
\label{app23}  
\end{equation}

In the following, we consider three limiting cases: photon propagation in vacuum, the LZ transition probability for a monotonic potential derived under the stationary-phase approximation, and the non-monotonic potential case where the LZ approximation ceases to be valid.
\subsection{Vacuum case:}
In the case of photon propagation in vacuum, where no plasma effects are present, the photon acquires no effective mass, i.e., $\delta(z)=0$ while $\delta_{A^\prime}$ is constant. Consequently, the phase becomes $\varphi(z^\prime)=-\delta_{A^\prime}(z^\prime-z_0)$. Therefore, we can write
\begin{equation}
\int ^z_{z_0}dz^\prime \delta_{A^\prime} e^{i\varphi(z^\prime)}=\frac{1-e^{-i\delta_{A^\prime}L}}{i}, 
\label{app24}
\end{equation}
with $L=z-z_0$. Thus, the active-sterile oscillation probability becomes
\begin{equation}
P_{a\leftrightarrow n}=4\epsilon^2\sin^2\Big(\frac{\delta_{A^\prime}L}{2}\Big)+\mathcal{O}(\epsilon^3).
\label{app25}
\end{equation}

For a constant background, the active-sterile oscillation probability beyond resonance can be obtained as
\begin{equation}
P_{a\leftrightarrow n}=\sin^2(2\theta_m)\sin^2\Big(\delta_m L\Big),    
\label{probaility_standard}
\end{equation}
where 
\begin{equation}
\tan2\theta_m=\frac{2\epsilon \delta_{A^\prime}}{\delta_{A^\prime}-\delta(z)}, \quad \delta_m=\sqrt{\epsilon^2\delta_{A^\prime}^2+\frac{(\delta_{A^\prime}-\delta(z))^2}{4}}.  
\label{const2}
\end{equation}
In the limit $\epsilon\ll 1$ and $\delta=0$, Eq. \ref{probaility_standard} reduces to Eq.~\ref{app25}.

\subsection{Stationary-phase approximation and Landau-Zener formula}
When the phase $\varphi(z)$ is highly oscillatory, the conversion probability integral can be evaluated using the stationary-phase approximation, as the probability varies slowly near the stationary points $z_n$ satisfying $\varphi'(z_n)=0$. These correspond to the resonance condition $\delta(z_n)=\delta_{A'}$. Expanding the phase around each stationary point yields
\begin{equation}
\varphi(z) \simeq \varphi(z_n) + \tfrac{1}{2}\varphi^{\prime\prime}(z_n)(z-z_n)^2,
\qquad
\delta_{A^\prime}(z) \simeq \delta_{A^\prime}(z_n),
\label{app26}
\end{equation}
and each saddle contributes
\begin{equation}
\int dz~ \delta_{A^\prime}(z)e^{i\varphi(z)} \simeq
\sum_n \delta_{A^\prime}(z_n)e^{i\varphi(z_n)}
\sqrt{\frac{2\pi}{|\varphi^{\prime\prime}(z_n)|}}e^{i\frac{\pi}{4}\sigma_n},
\label{app27}
\end{equation}
where $\sigma_n = \mathrm{sgn}[\varphi^{\prime\prime}(z_n)]$.

Therefore, the transition amplitude and probability become
\begin{align}
\mathcal{A}_{a\to n} &\simeq
\epsilon \sum_n
\sqrt{\frac{2\pi\delta_{A^\prime}^{2}(z_n)}{|\varphi^{\prime\prime}(z_n)|}}
e^{i[\varphi(z_n)+\frac{\pi}{4}\sigma_n]},
\label{app28} \\
P_{a\to n} &= |\mathcal{A}_{a\to n}|^2
= \epsilon^2\left[\sum_n A_n
+\sum_{n<m}2\sqrt{A_n A_m}\cos(\varphi_{nm})\right],
  \label{app29}
  \end{align}
  where the individual resonance contribution is defined as
  \begin{equation}
  A_n = \frac{2\pi\delta_{A'}^{2}(z_n)}{|\Phi^{\prime\prime}(z_n)|},
  \label{app30a}
  \end{equation}
  and the accumulated phase between two resonances is
  \begin{equation}
  \varphi_{nm}=\varphi(z_n)-\varphi(z_m)+\tfrac{\pi}{4}(\sigma_n-\sigma_m)
  \simeq \int_{z_m}^{z_n}dz(\delta(z)-\delta_{A^\prime}).
  \label{app31}
  \end{equation}
  When the phase difference satisfies $\varphi_{nm}\gg 2\pi$ (corresponding to many oscillations within the observational bandwidth or line-of-sight variations), the interference term averages to zero, and the probability reduces to the incoherent sum \cite{Landau:1932vnv,Zener:1932ws}
  \begin{equation}
  P_{a\to s}\simeq \epsilon^2 \sum_n A_n,
  \label{app32}
  \end{equation}
  which corresponds to the non-adiabatic LZ limit, where single-resonance probabilities add incoherently and phase effects are neglected.

This stationary-phase treatment remains valid as long as the background potential varies monotonically near each resonance. In the next subsection, we discuss the breakdown of this approximation for non-monotonic potentials, where the resonant conversion occurs close to a local extremum of the in-medium photon mass and the standard LZ description no longer applies.

\subsection{Breakdown of the Landau-Zener approximation for non-monotonic potential}

When the DP mass lies near a local extremum of the in-medium photon mass, $m_{\gamma}(z)$, two nearby resonances $z_1$ and $z_2$ appear and approach one another as $m_{A^\prime}\to m_{\gamma}(z_{\mathrm{crit}})$. At the critical mass
\begin{equation}
m_{A^\prime} = m_{\gamma}(z_\mathrm{crit}) \equiv m_{\mathrm{crit}},
\end{equation}
the two resonances coalesce at the extremum $z_\mathrm{crit}$. Near this point, the first two derivatives of the phase vanish,
\begin{equation}
\varphi^\prime(z_{\mathrm{crit}})=\varphi^{\prime\prime}(z_{\mathrm{crit}})=0,\qquad \varphi^{\prime\prime\prime}(z_{\mathrm{crit}})\neq 0,
\end{equation}
and the usual single-saddle LZ treatment fails. The departure from the LZ regime is characterized by
\begin{equation}
\xi=\frac{\varphi^{\prime\prime}(z_n)}{|\varphi^{\prime\prime\prime}(z_{\mathrm{crit}})|^{2/3}},
\label{app33}
\end{equation}
with $\xi\lesssim 1$ indicating the breakdown region.

Expanding around $z_\mathrm{crit}$, the phase and diagonal terms become
\begin{align}
\varphi(z)&\simeq \varphi(z_\mathrm{crit})+\tfrac{1}{3!}\varphi^{\prime\prime\prime}(z_\mathrm{crit})(z-z_\mathrm{crit})^3,
\label{app34}\\
\delta_{A^\prime}(z)&\simeq \delta_{A^\prime}(z_{\mathrm{crit}})+\delta'_{A^\prime}(z_{\mathrm{crit}})(z-z_\mathrm{crit}),
\label{app35}
\end{align}
and introducing the Airy rescaling
\begin{equation}
u=\sigma_c\left(\frac{\varphi^{\prime\prime\prime}(z_{\mathrm{crit}})}{2}\right)^{1/3}(z-z_{\mathrm{crit}}),\qquad
dz=\left(\frac{2}{|\varphi^{\prime\prime\prime}|}\right)^{1/3}du.
\label{app36}
\end{equation}
Following Eq.~\ref{app22}, the transition probability becomes
\begin{equation}
P_{a\to n}\simeq \epsilon^2
\left|
\left(\frac{2}{|\varphi^{\prime\prime\prime}|}\right)^{1/3}\int du
\Big[\delta_{A^\prime}(z_\mathrm{crit})+\sigma_c\left(\frac{2}{|\varphi^{\prime\prime\prime}|}\right)^{1/3}u \delta^\prime_{A^\prime}(z_{\mathrm{crit}})\Big]
e^{i(\zeta u+\frac{u^3}{3})}
\right|^2,
\label{app37}
\end{equation}
where $\zeta=\sigma_c(2/|\varphi^{\prime\prime\prime}|)^{1/3}\varphi'$. Using the standard Airy integrals
\begin{equation}
2\pi Ai(\zeta)=\int_{-\infty}^\infty due^{i(u^3/3+\zeta u)},\qquad
2\pi Ai^\prime(\zeta)=\int_{-\infty}^\infty duue^{i(u^3/3+\zeta u)},
\end{equation}
the compact result for the conversion probability is obtained as \cite{Brahma:2023zcw}
\begin{equation}
P_{a\leftrightarrow n}\simeq
4\pi^2\epsilon^2\delta_{A'}^2
\left(\frac{2}{|\varphi^{\prime\prime\prime}|}\right)^{2/3}
\left[
Ai(-\zeta)
+i\sigma_c\left(\frac{2}{|\varphi^{\prime\prime\prime}|}\right)^{1/3}
\left(\frac{\delta'_{A'}}{\delta_{A'}}-\frac{\varphi^{\prime\prime\prime\prime}}{6\varphi^{\prime\prime\prime}}\right)
Ai^\prime(-\zeta)
\right]^2_{z=z_\mathrm{crit}}.
\label{app39}
\end{equation}

Equation.~\ref{app39} applies in the vicinity of the critical point, $(m_{A^\prime}\simeq m_{\mathrm{crit}})$, where $\xi\lesssim 1$ and the standard LZ approximation no longer holds.

\section{Photon-dark photon oscillation in Messier 87 supermassive black hole background}\label{sec3}

In this section, we explore three representative electron-density models adopted for the environment around the M87 BH and describe their corresponding plasma profiles. Using these density models, we analyze the variation in the probability of photon-DP conversion with the oscillation length scale and the deviation of the DP mass from its critical value. The behavior of the oscillation probability is explored for different choices of the electron density profile and critical mass parameter, considering both monotonic and non-monotonic plasma potentials.

\subsection{Electron Density Profiles Around M87*}

The plasma surrounding the SMBH M87* is often modeled using simplified accretion prescriptions. A standard choice is the spherical (Bondi-like) accretion model, which assumes a steady, spherically symmetric inflow. Imposing a constant mass accretion rate,
\begin{equation}
\dot{M}=4\pi r^{2}\rho v_r ,
\end{equation}
and adopting a free-fall velocity profile $v_r\propto r^{-1/2}$, the density scales as $\rho\propto r^{-3/2}$. The corresponding electron number density follows a monotonic power-law profile \cite{Rybicki1979RadiativePI,Quataert:2002xn,Yuan:2014gma,EventHorizonTelescope:2019pgp,Roy:2023rjk,Hada:2024icg}
\begin{equation}\label{plasma_power_law}
n_e(r)=n_0\left(\frac{r}{r_{\rm ph}}\right)^{-3/2},
\end{equation}
where $n_0$ denotes the electron density near the photon sphere with radius $r_{\rm ph}$. For M87*, the photon sphere radius is approximately $r_{\rm ph} \simeq 6.2\times10^{-7}\,\mathrm{kpc}$ from the center of the SMBH.

The radiative accretion flow models \cite{Yuan:2014gma} predict that the plasma in the vicinity of the event horizon is hot and magnetized, with electron temperatures $T_e\sim10^9\ \mathrm{K}$ and magnetic fields of order $(1\text{–}30)\ {\rm G}$. For M87, mm-wavelength observations imply electron densities $n_e\sim10^5\text{–}10^7\ {\rm cm^{-3}}$ at radii $r\simeq(5\text{–}10)~r_g$ (where $r_g=GM$ is the gravitational radius), are consistent with sub-Eddington accretion rates of $\dot{M}\sim10^{-5}~M_\odot~\mathrm{yr^{-1}}$. These values correspond to a dilute, optically thin, magnetically dominated plasma environment. Since, the magnetic field is relatively weak, QED-induced photon mass corrections are negligible, and the effective photon mass is instead governed by the plasma frequency,
\begin{equation}
m_{\rm eff}(r)\simeq \omega_{\rm pl}(r)=\sqrt{\frac{4\pi\alpha n_e(r)}{m_e}},
\label{nd3}
\end{equation}
so that resonant photon-DP mixing is dictated by the interplay between the plasma mass and the DP mass. Here, in Eq. \ref{nd3}, $\alpha$ denotes the fine structure constant and $m_e$ denotes the electron mass. For our analysis, we adopt a reference electron number density $n_0 = 10^{6}\,\mathrm{cm^{-3}}$, motivated by accretion based modeling studies of the environment surrounding M87*, as discussed above.

We also consider a spherically symmetric plasma distribution motivated by accretion-disk simulations, which can be approximated by a log-normal profile \cite{Zhang:2022osx}
\begin{equation}\label{plasma_lognormal}
n_e(r)=n_{\max}\exp\left[-\frac{\left(\ln(r/r_m)\right)^{2}}{2\sigma^{2}}\right],
\end{equation}
where $r_m$ denotes the radius at which the density peaks, $\sigma$ determines the width of the distribution, and $n_{\max}$ is the maximum density. For illustration, we adopt $r_m =3 r_{\rm ph}$ and $n_{\max}\sim10^{8}~\mathrm{cm^{-3}}$, consistent with plasma densities required to generate radio emission at MHz frequencies in the inner flow. A representative value $\sigma=0.5$ yields a moderately narrow peak, producing two spatial locations along the line of sight where $m_{A^\prime} = m_{\rm eff}(r)$. These dual resonances naturally give rise to coherent interference between successive level crossings, enhancing the overall photon-DP conversion probability. This behavior is well described by the multiple level crossing formalism developed for non-monotonic plasma potentials as described in Section~\ref{sec2}.

To complement these near-horizon models, we additionally adopt free-electron density profiles from \cite{Ning:2024eky}, based on the Illustris-TNG300 cosmological magnetohydrodynamical simulations \cite{Marinacci:2017wew,Pillepich:2019bmb,Nelson:2019jkf}. The Illustris and its successor Illustris TNG projects aim to model the universe from shortly after the big bang to the present epoch, capturing the coupled dynamics of DM, baryons, and the SMBHs within the standard cosmological framework. TNG300 with a box size of $300~\rm Mpc$ provides an unparalleled statistical environment for studying massive galaxies and clusters, including M87-like systems. In Virgo-like cluster analogous with total masses $(6.3\pm0.9)\times10^{14}~\mathrm{M}_\odot$, the number density reaches $10^{-2} ~\rm{cm^{-3}}$ within the inner tens of kpc and remains at $10^{-3} ~\rm{cm^{-3}}$ on scales of several hundred kpc. Such an extended and coherent field structure may also provide a suitable environment for photon-DP conversion. This plasma profile as presented in \cite{Ning:2024eky}, decreases with distance and is treated as monotonic profile in the present study.

\subsection{Photon-dark photon conversion probabilities in different plasma profiles around M87*}
\begin{figure*}
   \centering
    \includegraphics[width=0.45\linewidth]{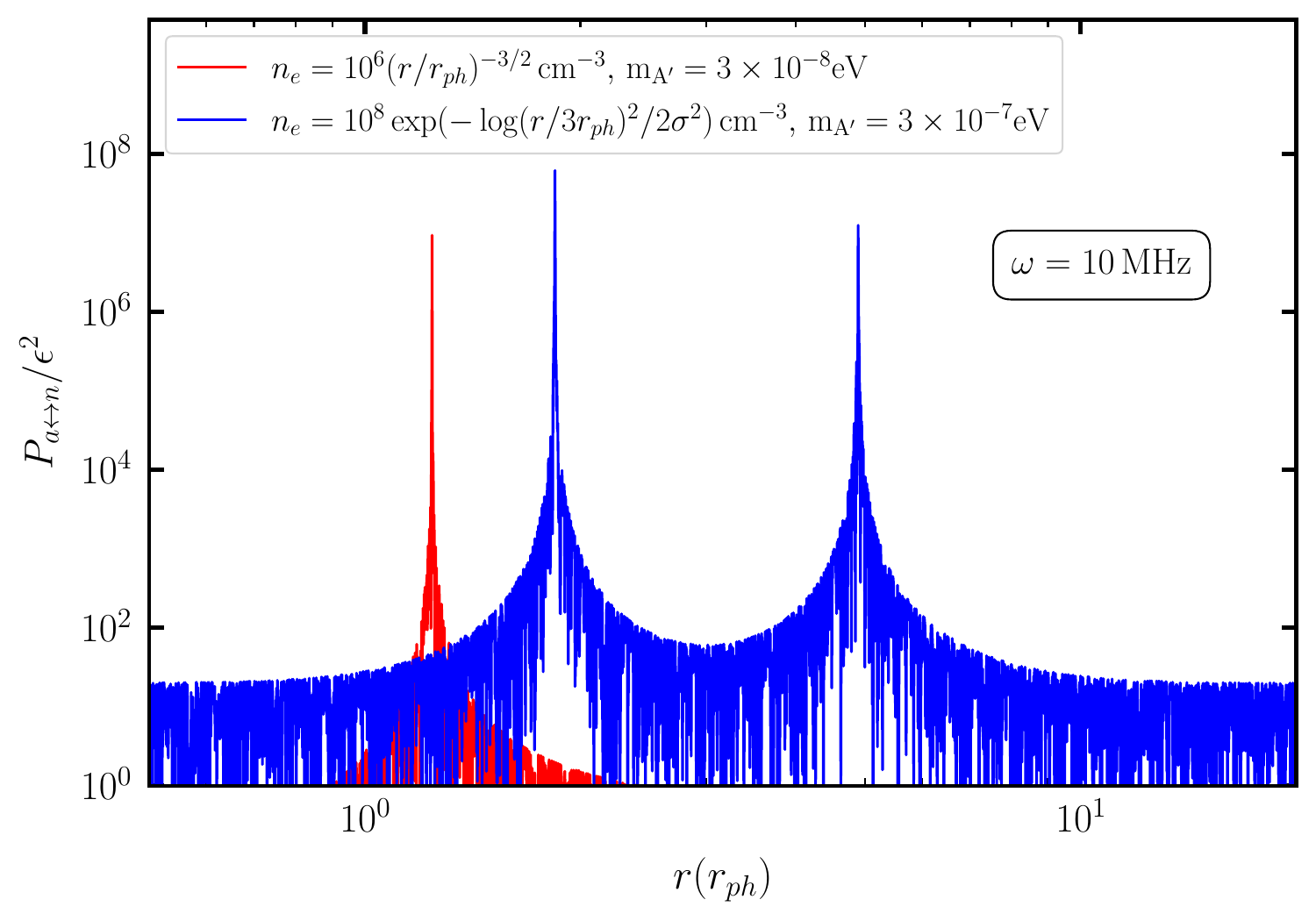}
   \includegraphics[width=0.45\linewidth]{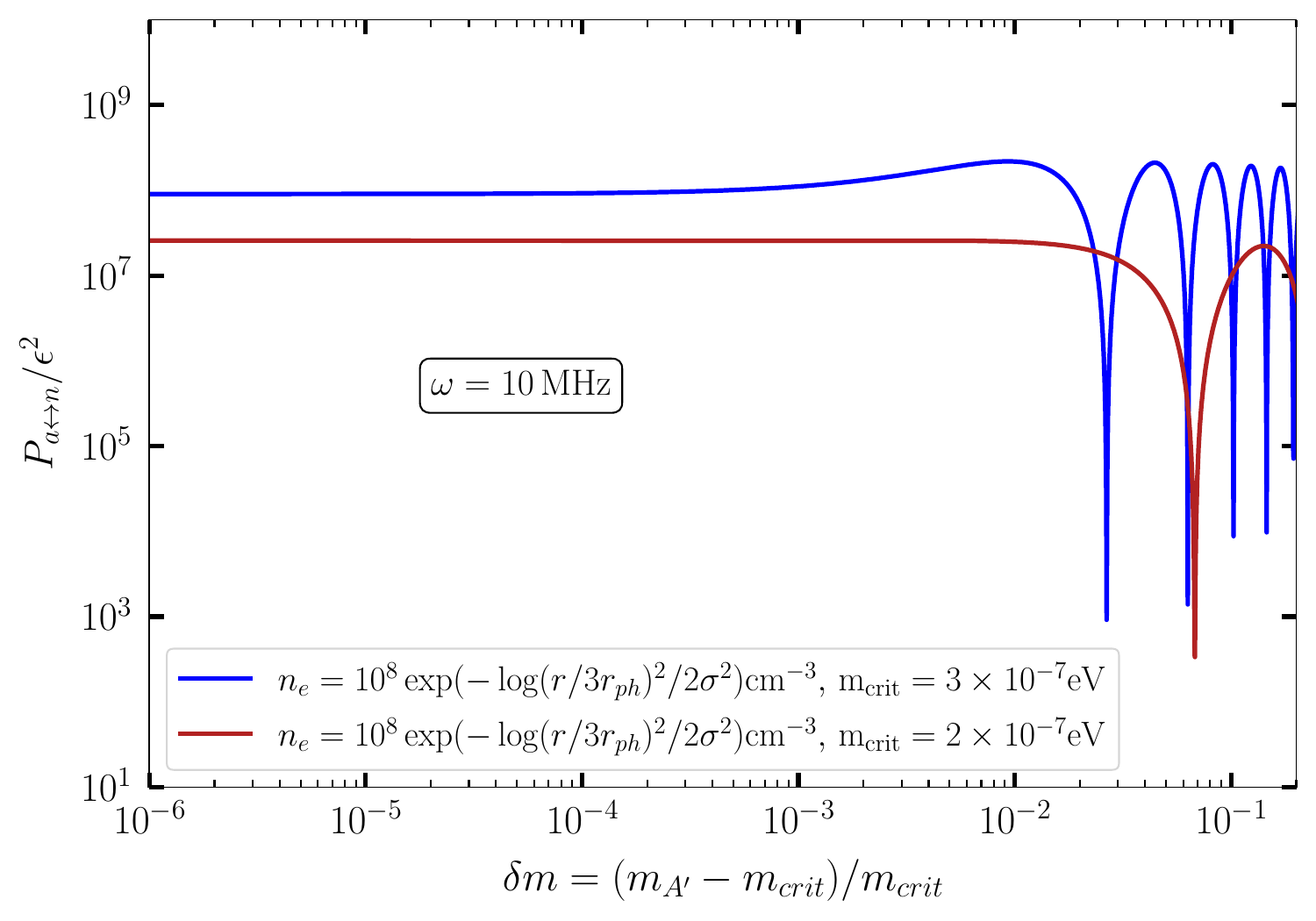}
   \caption{
Conversion probability $P_{a\leftrightarrow n}/\epsilon^{2}$ for photon-DP 
mixing.  
\textit{Left Panel:} Probability as a function of distance for a monotonic power-law 
plasma profile (red) showing a single resonance near the photon-sphere radius, and 
for a non-monotonic log-normal profile (blue) exhibiting two resonant locations.  
\textit{Right Panel:} Probability plotted against the mass-deviation parameter 
$\delta m$, computed using the 
multiple-level-crossing approximation Eq.~\ref{app39} and evaluated at the plasma peak $r_m=z_{c}$.}
    \label{fig:smbhprob}
\end{figure*}

In this section, we present the conversion probability between DPs and 
ordinary photons.  Using the power-law plasma profile Eq.~\ref{plasma_power_law}, which enters in  $\delta(z)$, the transition probability $P_{a\leftrightarrow n}$ is obtained from the standard expression Eq.~\ref{probaility_standard}. For a monotonic plasma profile, the probability attains its maximum value at the distance $L$ for which $\delta(z)=\delta_{A^\prime}$. In left panel of FIG.~\ref{fig:smbhprob} we show the probability, $P_{a\leftrightarrow n}/\epsilon^{2}$, as a function of distance (red curve), evaluated for a DP mass $m_{A'} = 3\times 10^{-8}\,\mathrm{eV}$ and a photon frequency of $10\,\mathrm{MHz}$. A sharp resonance appears near the photon-sphere radius ($r/r_{\rm ph}=1.5$) when the central number density is $n_{0}=10^{6}\,\mathrm{cm^{-3}}$.

For comparison, we also show the probability obtained for a non-monotonic, 
log-normal plasma profile Eq.~\ref{plasma_lognormal} (blue curve).  
As illustrated in left panel of FIG.~\ref{fig:smbhprob}, this profile yields two distinct resonant locations, 
reflecting the non-monotonic behavior of the plasma distribution.  
The plot corresponds to $m_{A^\prime} = 3\times 10^{-7}\,\mathrm{eV}$ and 
$n_{0}=10^{8}\,\mathrm{cm^{-3}}$, with the width parameter of the log-normal 
distribution fixed at $\sigma=0.5$, which produces a moderately peaked profile.

In the right panel of FIG.~\ref{fig:smbhprob}, we show $P_{a\leftrightarrow n}/\epsilon^{2}$ as a function of the 
DP mass deviation relative to the critical mass 
$m_{\rm crit}$, i.e., $\delta m=(m_{A^\prime}-m_{\rm crit})/m_{\rm crit}$.  
This plot is obtained using the approximate expression applicable to systems with 
multiple level crossings Eq.~\ref{app39}, at fixed $m_{\rm crit}$ with the values $2\times 10^{-7}~\mathrm{eV}$ (red curve) and $3\times 10^{-7}~\mathrm{eV}$ (blue curve) and photon frequency 
$\omega$. These benchmark values of
the critical mass are selected to maximize the conversion probability for the given plasma density, frequencies around $10\,\mathrm{MHz}$, and an oscillation length of the order of $r_{\rm ph}$.
For the same non-montonic plasma profile adopted in left figure of FIG.~\ref{fig:smbhprob}, we compute the required 
derivatives $\phi'''$ and $\phi''''$ and evaluate the transition probability at 
$z=z_{c}$, the location of the plasma peak.  
This resonant distance $z_{c}$ corresponds to the value of $m_{\rm crit}$ for which 
the plasma number density is maximized. 

We do not show the DP-photon conversion probability computed using the Illustris-TNG plasma density profile for the SMBH environment in FIG.~\ref{fig:smbhprob} for the following reason. The plasma distribution extracted from the Illustris-TNG simulation characterizes the large-scale galactic medium, with typical length scales extending to kpc and beyond, whereas FIG.~\ref{fig:smbhprob} corresponds to plasma profiles on much smaller, near-horizon scales. While the conversion probability is not displayed, the corresponding constraints on the coupling are derived using the Illustris-TNG plasma profile and presented in Section.~\ref{sec_limits}.

\section{Photon-dark photon oscillation in Crab Nebula background}\label{sec4}
%In this section, we examine the non-monotonic behavior of the photon effective mass in the Crab Nebula environment.

In this section, we examine the non-monotonic behavior of the effective mass relevant for photons associated with the Crab Nebula emission. The Crab Nebula is a pulsar wind Nebula powered by the central Crab pulsar. The relativistic wind of the pulsar injects particles and energy into the Nebula and drives its observed radiation \cite{10.1093/mnras/167.1.1,Buhler:2013zrp}. The photons considered here originate near the pulsar and initially propagate through its magnetospheric plasma and
we therefore employ the standard Goldreich-Julian(GJ) model \cite{Goldreich:1969sb} for this NS magnetosphere, in which the GJ charge density determines the plasma-induced photon mass. We also analyze the resulting photon-DP conversion probability in the GJ background of Crab Nebula.

\subsection{Electron density profiles in Crab Nebula background}
The plasma environment in the Crab pulsar magnetosphere varies strongly with both radial and angular distances, leading to substantial changes in the plasma-induced photon mass as photons propagate through the magnetosphere. Rotation of the NS generates an induced electric field that exceeds the gravitational binding energy at the stellar surface. As a result, surface charges are stripped off and flow along magnetic field lines until they redistribute such that the local Lorentz force vanishes. This establishes a corotating magnetosphere whose electron number density is well described by the GJ prescription as
\begin{equation}
n_{e}^{\text{GJ}}(r,\theta) = \frac{2  \mathbf{\Omega
}\cdot\textbf{B}}{e}  \Big[\mathcal{F}_{1}({r_{\rm red}})  \sin^{2}\theta 
 -\mathcal{F}_{2}({r_{\rm red}}) \left(\sin^{2}\theta-2\cos^{2}\theta\right)\Big],  
 \label{GJ1}
\end{equation}
where the functions in the square brackets encode the relativistic corrections to the canonical GJ density, given as
\begin{equation}
\begin{split}
\mathcal{F}_{1}({r_{\rm red}})=r_{\rm red}^{3} \Bigg\{ \bigg(1-\frac{\kappa}{r_{\rm red}^{3}}\bigg)\bigg[ \frac{2}{r_{\rm red}-1}-\frac{1}{(r_{\rm red}-1)^{2}}
+2\ln\Big(1-\frac{1}{r_{\rm red}}\Big)\bigg] \Bigg\}+\\
 \bigg(2+\frac{\kappa}{r_{\rm red}^{3}}\bigg)\Bigg\{ \frac{1}{r_{\rm red}}+\frac{1}{r_{\rm red}-1}+2\ln\Big(1-\frac{1}{r_{\rm red}}\Big)\Bigg\},   
\end{split}
\label{GJ2}
\end{equation}
and
\begin{equation}F_{2}({r_{\rm red}})={r_{\rm red}}^{3}\frac{2\left(1-\frac{\kappa}{{r_{\rm red}}^{3}}\right)}{1-\frac{1}{{r_{\rm red}}}}\Bigg\{ \frac{1}{2{r_{\rm red}}^{2}}+\frac{1}{{r_{\rm red}}}+\ln\Big(1-\frac{1}{{r_{\rm red}}}\Big)\Bigg\},
\label{GJ3}
\end{equation}
where $r_{\rm red}=r/r_g$ and $\kappa=2/5(R/r_g)^2$ with $R$ denotes the radius of the NS, $\Omega$ denotes its angular velocity, $B$ is the magnetic field and $e$ stands for the electric charge. Considering the magnetic field is dipolar outside the star and the magnetic moment axis is tilted by an angle $\alpha_m$ with the rotation axis, which is necessary for the pulsed radiation, and also consider that the angular velocity is along the $z$ axis, Eq. \ref{GJ1} becomes
\begin{equation}
n_{e}^{\text{GJ}}(r,\theta)=\frac{B_{0}}{eR_{\mathrm{LC}}} \left(\frac{R}{r}\right)^{3} \left(3\cos\theta \mathbf{\hat{m}}\cdot\mathbf{\hat{r}}-\cos\alpha_{m}\right)\Big[\mathcal{F}_{1}({r_{\rm red}})  \sin^{2}\theta -\mathcal{F}_{2}({r_{\rm red}}) \left(\sin^{2}\theta-2\cos^{2}\theta\right)\Big],
\label{GJ4}
\end{equation}
where $B_0$ denotes the surface magnetic field of the magnetized star, $R_{\mathrm{LC}}=1/\Omega$ denotes the light cylinder (LC) radius, $\mathbf{\hat{m}}$ denotes the direction of the magnetic moment axis and 
\begin{equation}
 \mathbf{\hat{m}} \cdot \mathbf{\hat{r}}=\cos\theta\cos\alpha_{m}+\sin\theta\sin\alpha_m \cos \left(\Omega t\right).
 \label{GJ5}
\end{equation}

The LC radius marks the distance out to which plasma in the magnetosphere can corotate with the NS. Since the corotating charges have tangential velocity $\mathbf{v}=\boldsymbol{\Omega}\times\mathbf{r}$, their speed increases linearly with radius. For a fixed angular velocity $\Omega$, causality requires that this velocity never exceed the speed of light. The point at which $v=1$ (speed of light in vacuum) defines the LC radius, yielding $R_{\rm LC}=1/\Omega$. In regions where the magnetic field strength satisfies $B(r)\gtrsim B_{ c}=m_e^2/e$, both the magnetic field and the plasma contribute to the effective photon mass, but with opposite signs. The total effective mass may be written as
\begin{equation}
m_\gamma^2 = V_{\rm QED} + V_{\rm pl},
\end{equation}
where the quantum electrodynamics (QED) contribution is
\begin{equation}
V_{\rm QED}\simeq -\frac{7\alpha}{45\pi} b_{\rm red}^2\eta \omega^2 \sin^2\theta ,
\label{GJ6}
\end{equation}
with $b_{\rm red}=B(r)/B_{c}$, where the radial component of the magnetic field, $B(r)$, falls off as $1/r^3$ outside the NS. The function $\eta$ encodes the leading behavior in both the weak-and strong-field limits and is well approximated by
\begin{equation}
\eta = \frac{1+1.2 b_{\rm red}}{1+1.33 b_{\rm red}+0.56 b_{\rm red}^2}.
\label{GJ7}
\end{equation}
The term $V_{\rm QED}$ arises from standard quantum electrodynamical effects such as vacuum birefringence, photon splitting, and magnetic pair production. The plasma contribution, sourced by the GJ charge density, takes the form
\begin{equation}
V_{\rm pl} = \omega_{\rm pl}^2 \sin^2\theta ,
\label{GJ8}
\end{equation}
where $\omega_{\rm pl}^2 = 4\pi\alpha n_e^{\rm GJ}/m_e$ is the plasma frequency with $\alpha$ the fine structure constant and $m_e$ denotes the mass of the electron. Because $V_{\rm QED}$ and $V_{\rm pl}$ carry opposite signs, their competition naturally produces a non-monotonic radial dependence of the effective photon mass, independent of the specific relativistic correction functions.

The radial behavior of the two potential terms differs significantly. The plasma contribution scales as $1/r^{3}$, whereas the QED term falls off more rapidly as $1/r^{6}$. At intermediate radii, where $V_{\rm QED}$ dominates, the effective photon mass becomes negative, $m_\gamma^{2}<0$, owing to the negative sign of the QED potential. Conversely, at larger radii the plasma term prevails and $m_\gamma^{2}>0$. Because the two contributions have opposite signs, there is a turnover point denoted the critical radius at which
\begin{equation}
V_{\rm pl}(r_{\rm crit}) \simeq - V_{\rm QED}(r_{\rm crit}) .
\end{equation}
This condition yields
\begin{equation}
r_{\rm crit} \sim (\omega^{2} R_{\rm LC} B_{0})^{1/3}\frac{eR}{m_e},
\label{GJ9}
\end{equation}
and defines an associated critical mass,
\begin{equation}
m_{\rm crit} \sim \frac{m_e}{e\omega R_{\rm LC}} .
\label{GJ10}
\end{equation}

The non-monotonic nature of $m_\gamma^{2}(r)$ implies that when the DP mass lies near $m_{\rm crit}$, two resonance points appear in close proximity, one near the stellar surface $R$ and the other near the LC radius $R_{\rm LC}$. Requiring the critical radius to fall within the interval $R \lesssim r_{\rm crit} \lesssim R_{\rm LC}$ leads to the condition
\begin{equation}
\frac{1}{R_{\rm LC}}
\lesssim
\frac{\omega^{2} B_{0} e^{2}}{B_{c} m_e}
\lesssim
\frac{R_{\rm LC}^{2}}{R^{3}} .
\label{GJ11}
\end{equation}

\subsection{Photon-dark photon conversion probabilities in the plasma of Crab Nebula}

\begin{figure*}
    \includegraphics[width=0.9\linewidth]{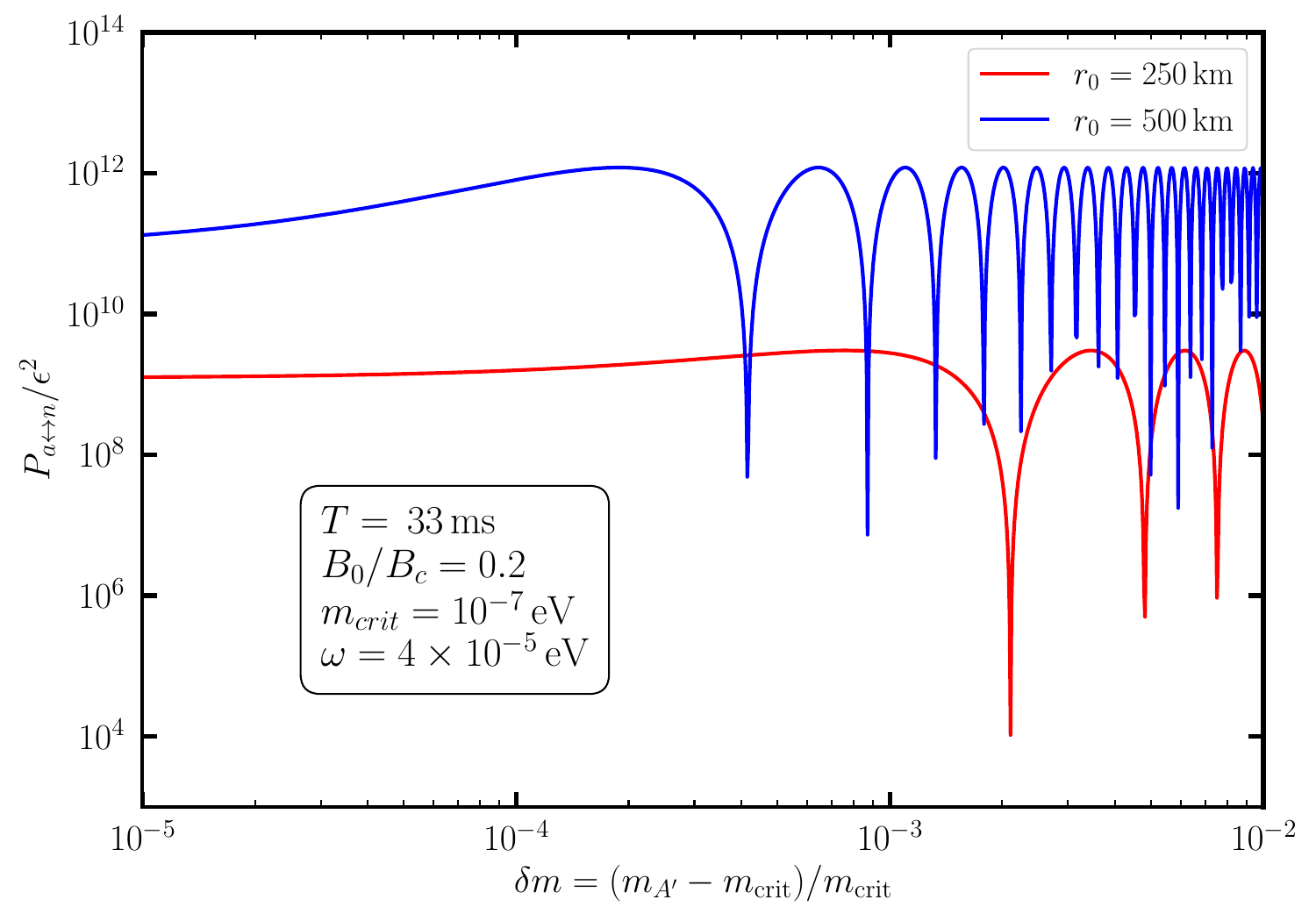}
    \caption{DP-photon conversion probability $P_{a\leftrightarrow n}$ as a function of the deviation parameter $\delta m$ in the Crab pulsar magnetosphere. The curves correspond to oscillation distances of $r_{0}=250\,\mathrm{km}$ (red) and $r_{0}=500\,\mathrm{km}$ (blue), computed for $m_{\rm crit}=10^{-7}\,\mathrm{eV}$ and a photon energy of $4\times10^{-5}\,\mathrm{eV}$. The pronounced oscillatory structure originates from the non-monotonic magnetospheric potential, whereas in the $\delta m \ll 1$ regime, the probability approaches a constant value, reflecting the need for the approximate expression Eq. \ref{app39} in the multiple-level crossing scenario.}
\label{fig:prob_crab}
\end{figure*}

In FIG.~\ref{fig:prob_crab} we present the photon-DP conversion probability 
$P_{a\leftrightarrow n}$ as a function of the deviation parameter $\delta m$ in the 
magnetospheric environment of the Crab pulsar.  
The curves are obtained by fixing the critical mass to 
$m_{\rm crit}=10^{-7}\,\mathrm{eV}$, the photon energy to 
$4\times 10^{-5}\,\mathrm{eV}$, and by considering two oscillation lengths,
$r_{0}=250\,\mathrm{km}$ and $r_{0}=500\,\mathrm{km}$.  
For the Crab pulsar, we adopt the LC radius 
$R_{\rm LC}=T/2\pi \sim 1600 \,\, \rm km$, with rotation period $T=33\,\mathrm{ms}$, 
and a magnetic-field strength satisfying $B_{0}/B_{c}=0.2$, 
characteristic of this source.  

The red curve corresponds to $r_{0}=250\,\mathrm{km}$, while the blue curve shows 
the result for $r_{0}=500\,\mathrm{km}$.  
As shown, the peak conversion probability increases with the oscillation 
distance.  
At larger values of $\delta m$, the highly oscillatory behavior reflects the 
non-monotonic structure of the pulsar magnetospheric potential entering through 
$\delta(z)$.  
In the regime $\delta m \ll 1$, the probability approaches a constant value, 
and the approximate expression Eq. \ref{app39} is required to accurately capture the 
behavior near the multiple-level-crossing resonance.

\section{Spectral Modeling and Data Analysis}\label{SPEC_Modeling}
In this section, we model the observed spectra of M87 and the Crab Nebula using LOFAR measurements and broadband SED data, respectively. In both cases, the intrinsic emission is well described by a power-law spectrum. For illustrative purposes, we introduce photon-DP mixing with a representative choice of oscillation parameters and fit the resulting modified spectrum to the observational data while retaining the same underlying power law form. The procedure is outlined below. 

\subsection{LOFAR data and M87*:}

\begin{figure*}
  \centering
  \includegraphics[width=8cm, height=6cm]{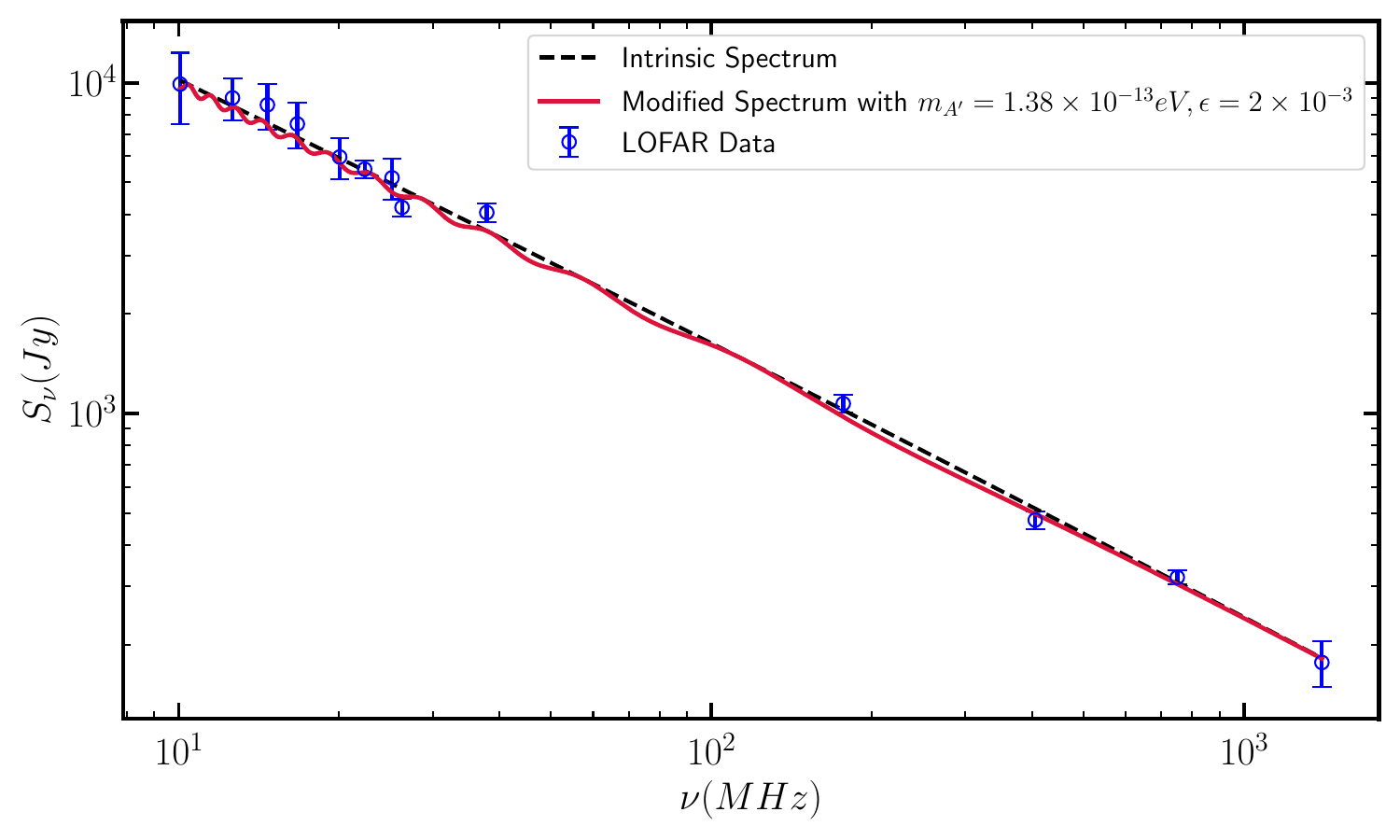}
  \includegraphics[width=8cm, height=6cm]{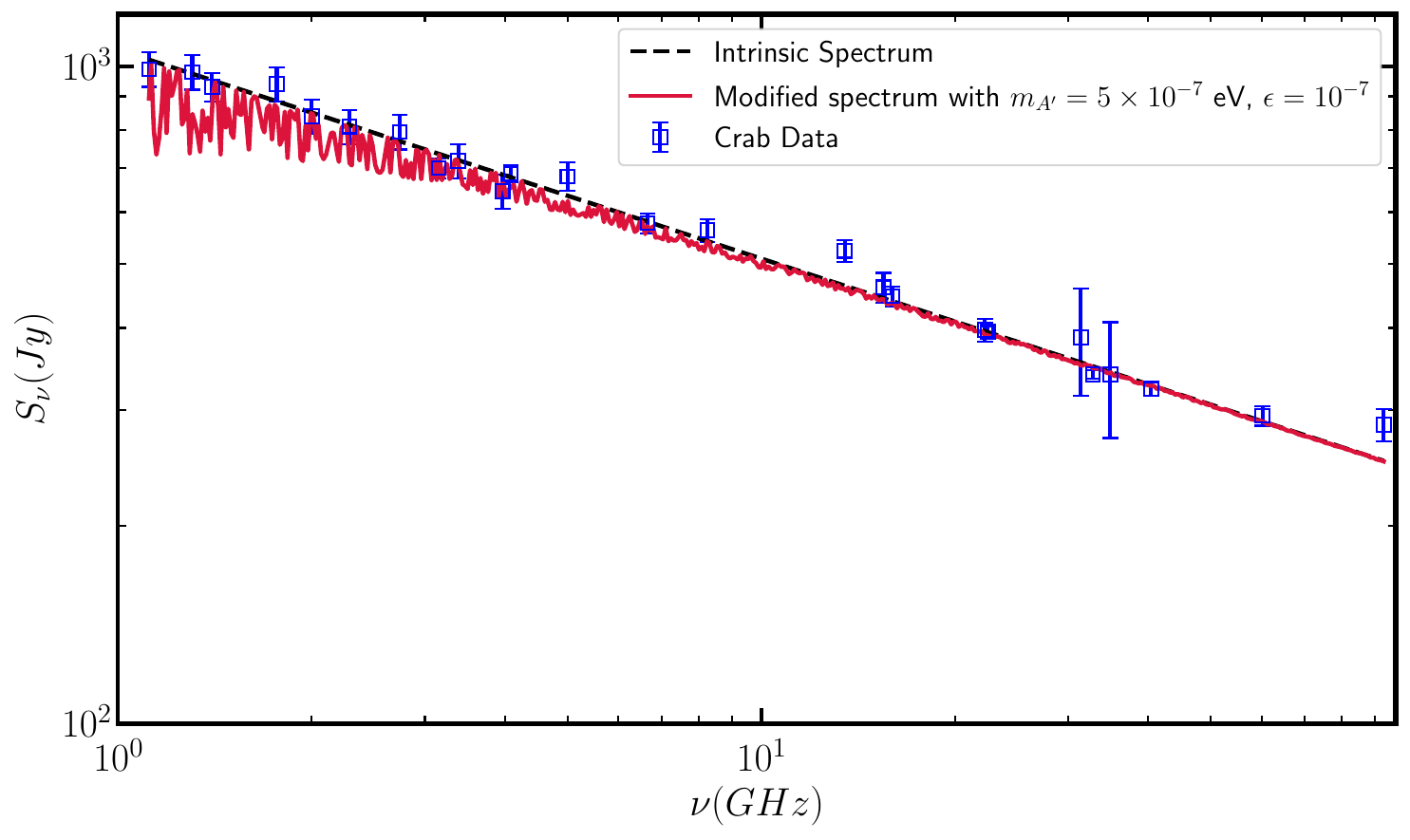}
  \caption{
\textit{Left panel:} Intrinsic and DP-modified radio spectrum of the M87* SMBH environment, using LOFAR data. 
The solid curve shows the distorted spectrum for the benchmark parameters 
$m_{A'} = 1.38\times10^{-13}\,\mathrm{eV}$ and $\epsilon = 2\times10^{-3}$.
\textit{Right Panel:} Intrinsic and DP-modified spectrum of the Crab Nebula. 
The red curve corresponds to the benchmark point 
$m_{A'} = 5\times10^{-7}\,\mathrm{eV}$ and $\epsilon = 10^{-7}$. 
Both panels illustrate how DP-photon conversion alters the radio continuum relative 
to the intrinsic emission models.
}
  \label{fig:spectral-fillting}
  \end{figure*}

We make use of low-frequency radio observations of the M87 galaxy obtained with the LOFAR array, covering the $10-1000$ MHz band, as indicated by the blue color bar in the left panel of FIG.~\ref{fig:spectral-fillting}. The central radio source associated with M87 is commonly identified as Virgo A, for which there is an extensive archival data and numerous literatures have modeled it \cite{Braude1969TheSO,Bridle1968OBSERVATIONSOR,Roger1969SpectralFD,Viner1975263MHzRS,1969ApJ...157....1K,1990PKS...C......0W,1980MNRAS.190..903L,2012MNRAS.423L..30S,2012A&A...547A..56D}.

For our analysis, we adopt the intrinsic flux-density model introduced in \cite{2012A&A...547A..56D}, which provides an excellent fit to the LOFAR data when expressed in logarithmic frequency space. The intrinsic spectrum is parametrized as
\begin{equation}\label{LOFAR_spec}
    S_{\mathrm{in}}(\nu) = A_0 \times 10^{\,A_1 \log_{10}\!\left(\frac{\nu}{150\,\mathrm{MHz}}\right) + A_2 \left[\log_{10}\!\left(\frac{\nu}{150\,\mathrm{MHz}}\right)\right]^2}\,\mathrm{Jy},
\end{equation}
  
where $\nu$ denotes the observing frequency. The best-fit parameters are $A_0 = 1.17\times10^{3}~\mathrm{Jy}$, where Jy stands for ``Jansky", used as the unit of flux,  $A_1=-0.81$, and $A_2=-0.01$ yield an accurate representation of the intrinsic radio emission arising from the plasma near the central SMBH of M87 which is shown in black dashed curve in the left panel of FIG.~\ref{fig:spectral-fillting}.

In addition to the intrinsic spectrum, we also show the modified flux predicted in the presence of 
photon-DP conversion. The modification arises from the frequency-dependent survival 
probability of photons propagating through the plasma environment surrounding the SMBH, which induces
a characteristic distortion of the radio continuum. The resulting spectrum, shown as a solid red curve in the left panel of FIG.~\ref{fig:spectral-fillting},
corresponds to a benchmark choice of DP parameters 
$m_{A'} = 1.38\times10^{-13}\,\mathrm{eV}$ and $\epsilon = 2\times10^{-3}$. 
This illustrates how the expected deviation from the intrinsic spectrum varies across the LOFAR band and 
highlights the frequencies at which the conversion effect produces the strongest suppression. 

\subsection{SED data and Crab Nebula:}

The Crab Nebula has been extensively observed across a broad frequency range, from radio to sub-millimeter wavelengths \cite{2010ApJ...711..417M,2010A&A...523A...2M,2011ApJ...730L..15Y}. A comprehensive spectral energy distribution is analyzed in \cite{2010ApJ...711..417M}, where the emission below $\sim 100~\mathrm{GHz}$ is dominated by a single synchrotron component produced by relativistic electrons in the pulsar wind Nebula. No significant thermal or dust contribution is required in this regime, and the observed fluxes are well described by a power-law spectrum of the form $A \nu^{\beta}$.

For frequencies up to $\sim 100~\mathrm{GHz}$, the best-fit spectral index is $\beta\simeq -0.3$, with no indication of a low-frequency turnover or additional curvature. This implies that processes such as free-free absorption or synchrotron aging contribute only marginally in the radio band, making this frequency range a reliable baseline for probing possible deviations from the intrinsic synchrotron emission, including photon-DP mixing effects.

Accordingly, in our analysis we model the intrinsic Crab Nebula spectrum with the power law fit to the low frequency data,
\begin{equation}\label{Crab_spec}
S_{\rm in}(\nu)=970\left(\frac{\nu}{1~\mathrm{GHz}}\right)^{-0.31} \mathrm{Jy},
\end{equation}
where the normalization and spectral index are obtained from the best fit to the observed radio fluxes.

The intrinsic power-law spectrum is indicated by the black dashed line in the right panel of FIG.~\ref{fig:spectral-fillting}. In addition to this, we also compute the modified flux expected in the 
presence of photon-DP conversion within the Crab pulsar magnetosphere. 
Because the effective mixing potential becomes non-monotonic due to the competing plasma and QED 
contributions, the resulting photon survival probability exhibits multiple level crossings and a 
characteristic frequency dependence. This induces a distortion in the observed radio spectrum whose 
magnitude and shape depend on the DP parameters and the assumed oscillation distance.

In right panel of FIG.~\ref{fig:spectral-fillting} we show, in red, the modified spectrum corresponding to the benchmark choice 
$m_{A'} = 5\times10^{-7}\,\mathrm{eV}$ and $\epsilon = 10^{-7}$. 
The deviation from the intrinsic synchrotron spectrum becomes most pronounced in the frequency bands 
where the conversion probability is enhanced by resonant or near-resonant mixing. 
This comparison illustrates how photon-DP oscillations may imprint detectable spectral 
signatures in the well-measured low-frequency emission of the Crab Nebula.

\section{Limits on photon-dark photon kinetic mixing}\label{sec_limits}

If DPs exist and are produced via photon-DP conversion in the plasma, they would imprint characteristic distortions on the observed spectrum. The resulting flux can be expressed as
\begin{equation}\label{th_flux}
S_{\rm th}(m_{A^\prime},\epsilon)
= S_{\rm in}\big[1 - P_{a\leftrightarrow n}(m_{A^\prime},\epsilon)\big].
\end{equation}
For the SMBH scenario, the intrinsic spectrum $S_{\mathrm{in}}$ follows Eq.~\ref{LOFAR_spec}, while for the Crab emission, it is given in Eq.~\ref{Crab_spec}.

To constrain these parameters, we compare the theoretical spectrum with the data using a chi-squared statistics,
\begin{equation}
\chi^{2}(m_{A^\prime},\epsilon)
= \sum_{i=1}^{N}
\left[
\frac{S_{\rm obs}^{i}-S_{\rm th}^{i}(m_{A'},\epsilon)}
{\sigma_{\rm obs}^{i}}
\right]^{2},
\end{equation}
In this expression, $S_{\rm obs}^{i}$ and $\sigma_{\rm obs}^{i}$ represent the observed flux and its corresponding uncertainty in the $i$-th frequency bin, whereas $S^{i}_{\mathrm{th}}$ is evaluated using Eq.~\ref{th_flux}.

While the least-squares framework allows for parameter estimation at arbitrary confidence level (C.L.), in this work we adopt the $95\%$ C.L., corresponding to $\Delta\chi^{2}=6.18$ for two degrees of freedom. 

\subsection{Constraints from Spectral Modeling of the M87*  Environment}

\begin{figure*}
    \includegraphics[width=0.9\linewidth]{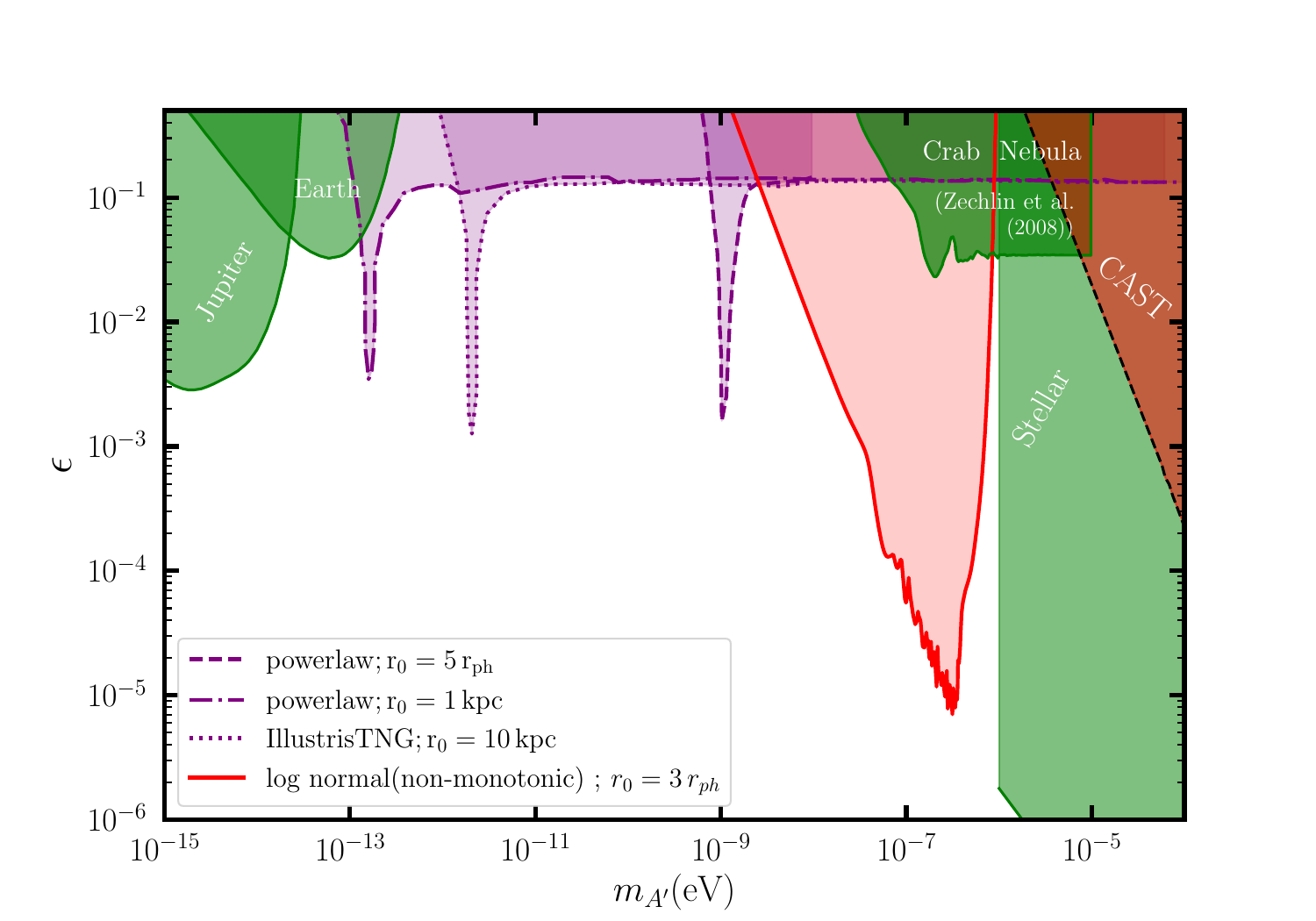}
    \caption{
Constraints on the kinetic mixing parameter $\epsilon$ as a function of the DP mass $m_{A'}$. 
The purple dashed and dot-dashed curves correspond to the cases where the DP-photon oscillation occurs near the photon sphere $(r_0 = 5\,r_{\rm ph})$ and at $1$ kpc scales, respectively, assuming a monotonic plasma profile. 
The dotted purple curve shows the limits obtained using the large-scale plasma density from the TNG300 simulation. 
The filled purple region indicates the parameter space excluded at the $95\%$ C.L. from the LOFAR radio observations. 
The red solid curve represents the constraint for a non-monotonic plasma profile near the SMBH, where multiple level crossings enhance the conversion probability. 
For comparison, other astrophysical and laboratory bounds are shown which are taken from the \texttt{AxionLimits} database \cite{AxionLimits}.}

    \label{fig:SMBH_limits}
\end{figure*}

With the intrinsic spectrum Eq. \ref{LOFAR_spec} of the emission near the SMBH M87* (see Section.~\ref{SPEC_Modeling}), we compare the
expected spectral distortion induced by DP to photon conversion with the observed LOFAR data.
This allows us to place constraints on the kinetic mixing parameter $\epsilon$ for fixed DP masses.

For the SMBH environment, we consider several plasma-density profiles.  
In the case of a monotonic power-law profile, the DP-photon conversion probability is evaluated for two
distinct oscillation-length scales:  (i) a region very close to the photon sphere, $r_0 = 5\,r_{\mathrm{ph}}$, and  
(ii) a much larger radius of order $\mathcal{O}(1\,\mathrm{kpc})$.  
The corresponding constraints are shown in FIG. \ref{fig:SMBH_limits} as dashed and dot-dashed purple curves, with the
LOFAR 95\% excluded region is represented by the filled band.  
We additionally incorporate a large-scale plasma-density estimate obtained from the TNG300 simulations,
whose resulting limits are plotted as the dotted purple curve.

As the figure illustrates, increasing the oscillation distance generally strengthens the bounds at lower
DP masses.  Conversely, for oscillation regions restricted to a few photon-sphere radii, the constraints tighten toward
higher DP masses. The sharp dips in the exclusion curves originates from the oscillatory phase of the DP-photon conversion probability. At specific values of the DP mass, the phase-matching condition, $(\delta(z)-\delta_{A^\prime})r_0 \sim \mathcal{O}(1)$ leads to constructive interference over the fixed propagation distance, maximizing the conversion probability. Consequently, a smaller mixing $\epsilon$ is sufficient to produce the same signal, resulting in a localized deep feature in the bound. 

Across the monotonic plasma profile scenarios, the strongest constraints reach at
$\epsilon \sim 10^{-3}$. For oscillation distances of
$r_0 = 5\,r_{\rm ph}$ and $r_0 = 1\,\mathrm{kpc}$ corresponding to the power law plasma profiles,
and $r_0 = 10\,\mathrm{kpc}$ for the IllustrisTNG profile,
the most stringent constraints are obtained at DP masses of
$1.03\times10^{-9}\,\mathrm{eV}$,
$1.13\times10^{-13}\,\mathrm{eV}$, and
$2.07\times10^{-12}\,\mathrm{eV}$, respectively.

We also investigate a non-monotonic plasma profile in Section.~\ref{sec3}, for which the resulting constraint is shown
as the solid red curve.  The presence of multiple density turning points near the SMBH induces multiple level crossings, enhancing the overall DP-photon conversion probability.  
Consequently, the resulting limits are significantly stronger than those from monotonic profiles.
For this case, the most stringent bound occurs at a DP mass of $m_{A'} = 5\times10^{-7}\,\mathrm{eV}$,
yielding a kinetic mixing constraint of $\epsilon \simeq 7\times10^{-6}$.

For completeness, we compare our constraints with existing astrophysical bounds obtained from
Jupiter \cite{Yan:2023kdg}, Earth-based atmospheric observations \cite{Fischbach:1994ir}, and nebular environments~\footnote{The bound reported by Zechlin et al. \cite{Zechlin:2008tj} is obtained from very-high-energy ($\mathcal{O}(\mathrm{TeV})$) gamma-ray observations of the Crab Nebula with Cherenkov telescopes. In that analysis, the propagation effects associated with the ambient plasma profile were not considered.}, where similar conversion
processes may occur. We additionally display the combined stellar-evolution limits and those from
CAST \cite{Redondo:2008aa}. Laboratory bounds are not included in the figure, as our focus here is on astrophysical
constraints in the relevant mass window.

While laboratory experiments provide strong bounds in this mass range, our analysis offers
competitive constraints from astrophysical environments, thereby extending the sensitivity
of indirect searches beyond terrestrial setups. In particular, within this mass regime our results
represent competitive to the available astrophysical constraints, highlighting the complementary role
of galaxy- and SMBH-based probes. Moreover, the limits derived here are largely independent of the
assumption that DP constitute the entirety of the DM, relying instead on
conversion processes governed by environmental plasma properties.

\subsection{Constraints from Spectral Modeling of Crab Nebula}

\begin{figure*}
    \centering
    \includegraphics[height=6cm,width=8.1cm]{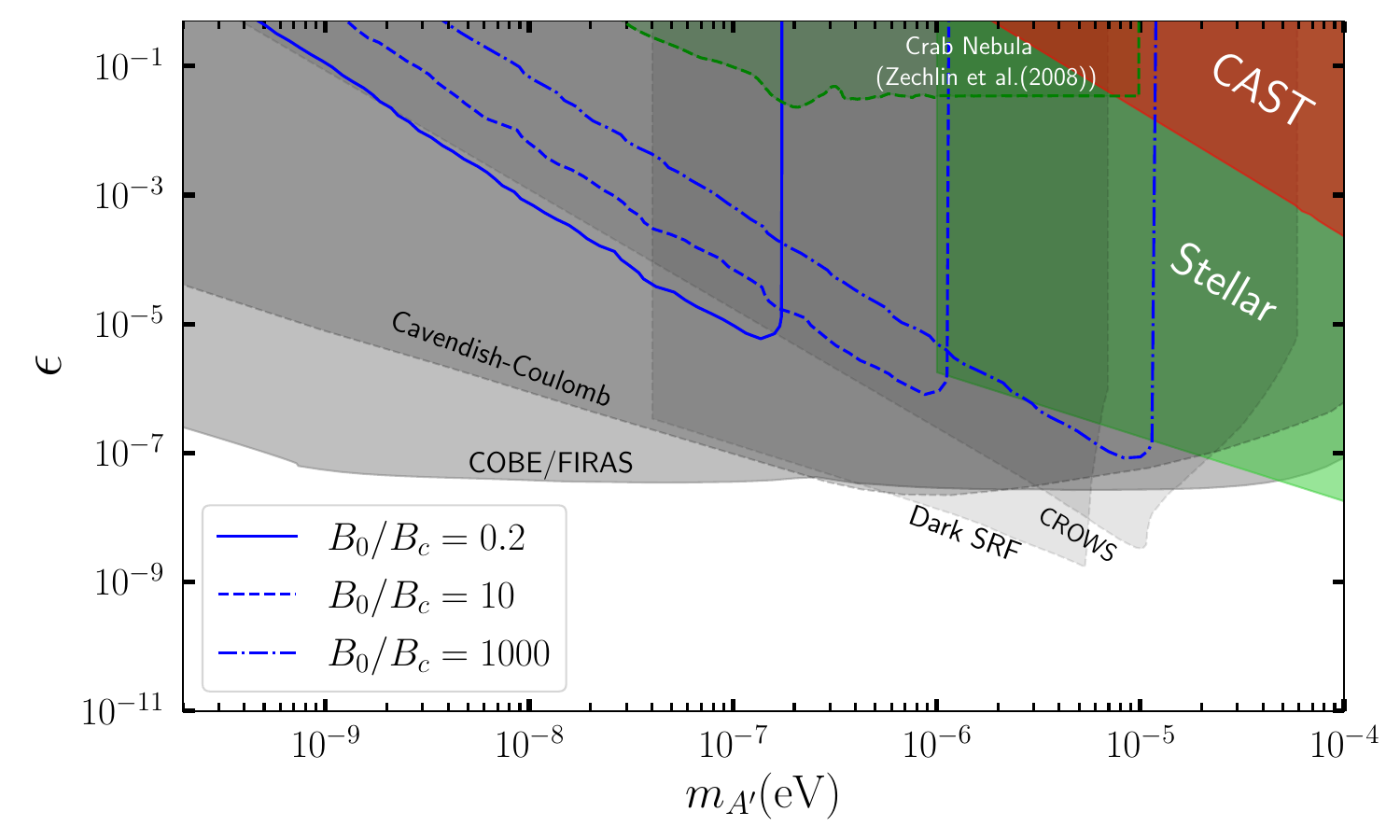}
    \includegraphics[height=6cm,width=8.1cm]{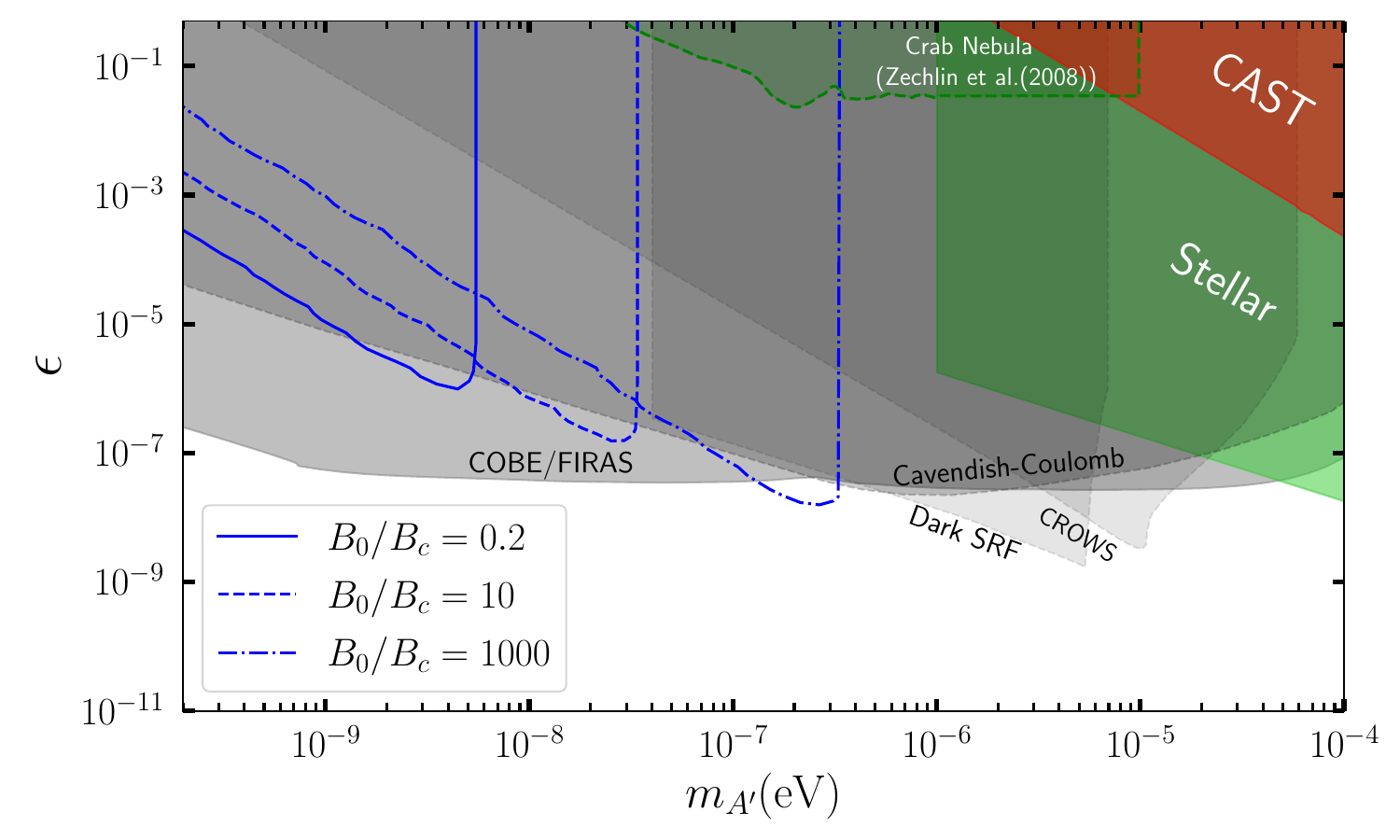}
    \caption{
    Constraints on the kinetic mixing parameter $\epsilon$ from the Crab pulsar: \textit{Left panel:} The $95\%$ exclusion limits for an oscillation distance of 
    $z = 200\,\mathrm{km}$. \textit{Right panel:} The $95\%$ exclusion limits for an oscillation distance of $z = 1000\,\mathrm{km}$. 
    The solid blue curves represent the bounds obtained using the non-monotonic effective potential 
    generated by the combined plasma and QED contributions in the Crab magnetosphere. 
    Dashed and dot-dashed curves denote projected sensitivities for enhanced magnetic field strengths 
    $B_0 = 10\,B_c$ and $B_0 = 1000\,B_c$, respectively.  
    Existing astrophysical limits from stellar cooling and CAST, very high energy gamma-ray observations from Crab Nebula, together with laboratory constraints
    %from conversion-based searches such as Dark SRF, Cavendish–Coulomb, and CROWS,
    as well as cosmological bounds from COBE/FIRAS, are included for comparison and are taken from the \texttt{AxionLimits} database~\cite{AxionLimits}.}
    \label{fig:Crab_constraints}
\end{figure*}

In the case of the Crab spectrum, we perform an analogous analysis to derive exclusion limits in the 
$\epsilon$-$m_{A'}$ parameter space. Owing to the fact that the plasma contribution and the QED vacuum 
polarization term enter the effective potential with opposite signs, the resulting profile becomes 
non-monotonic. Consequently, the conversion probability of DPs into photons must be evaluated 
using the formalism appropriate for multiple level crossings. The resulting $95\%$ exclusion limits are 
displayed as solid blue curves in both panel of FIG.~\ref{fig:Crab_constraints}.

For the Crab pulsar, we assume a surface magnetic field strength $B_0 = 0.2\,B_c$, where $B_c$, is the 
critical magnetic field. The rotation period is taken to be $T = 33\,\mathrm{ms}$, implying a 
LC radius $R_{\rm LC} = T/(2\pi)$. We consider two representative choices of the oscillation 
distance. For an oscillation length of $200\,\mathrm{km}$, the corresponding limits appear in left panel of FIG.~\ref{fig:Crab_constraints}, 
while the bounds obtained for an oscillation length of $1000\,\mathrm{km}$ are shown in the right panel of  FIG.~\ref{fig:Crab_constraints}.

We additionally display projected sensitivities for enhanced magnetic field strengths, with $B_0 = 10\,B_c$ 
and $B_0 = 1000\,B_c$, indicated by dashed and dot-dashed blue curves, respectively. The strongest 
constraint for the Crab pulsar in the $200\,\mathrm{km}$ scenario occurs at 
$\epsilon = 6\times 10^{-6}$ for a DP mass 
$m_{A'} \simeq 1.3\times 10^{-7}\,\mathrm{eV}$. In the $1000\,\mathrm{km}$ case, the tightest bound is 
$\epsilon \simeq 8\times 10^{-7}$ at $m_{A'} \simeq 4\times 10^{-9}\,\mathrm{eV}$. These sharp features arise 
in regions where the QED correction competes with the plasma contribution of the pulsar magnetosphere, 
leading to maximal conversion.

The strongest sensitivities in the $m_{A'}$-$\epsilon$ plane are obtained for 
$B_0/B_c = 10$ and $B_0/B_c = 1000$. 
For an oscillation distance of $200~\mathrm{km}$, the optimal sensitivities occur at 
$(m_{A'},\epsilon) = (8.70\times10^{-7}~\mathrm{eV},\,8.08\times10^{-7})$ and 
$(8.32\times10^{-6}~\mathrm{eV},\,8.3\times10^{-6})$, respectively, as indicated through the dashed and dot-dashed curves in the left panel of FIG.~\ref{fig:Crab_constraints}. As evident from the figure, the projected sensitivity is competitive with existing stellar bounds in this mass range.

For a larger oscillation distance of $1000~\mathrm{km}$, the strongest projected sensitivities shift to 
$(2.53\times10^{-8}~\mathrm{eV},\,1.54\times10^{-7})$ for $B_0/B_c = 10$ and 
$(2.66\times10^{-7}~\mathrm{eV},\,1.57\times10^{-8})$ for $B_0/B_c = 1000$, as shown in the dashed and dot-dashed curves in the right panel of FIG.~\ref{fig:Crab_constraints}.

For comparison, we also show existing astrophysical constraints from stellar cooling \cite{Dolan:2023cjs,Li:2023vpv,2015JCAP...10..015V,Schwarz:2015lqa}, very high energy gamma-ray observations from Crab Nebula \cite{Zechlin:2008tj} and CAST \cite{Redondo:2008aa}, together with laboratory bounds (Dark SRF \cite{Romanenko:2023irv}, Cavendish-Coulomb \cite{Kroff:2020zhp}, CROWS \cite{Betz:2013dza}), including those from cosmological COBE/FIRAS bounds \cite{Caputo:2020bdy,McDermott:2019lch}, which are indicated in gray. Notably, projected sensitivities for stronger magnetic fields 
while keeping the pulsar radius and rotation period fixed can approach, and in some regimes surpass, the 
current laboratory bounds, particularly when the oscillation distance is large. For $B_0/B_c = 1000$, the projected limits approach and can challenge the strongest existing laboratory bounds. Overall, the strongest constraints arise at higher DP masses when the oscillation distance is short, whereas for lighter 
masses the limits tighten as the oscillation distance increases.

\section{Conclusions and discussions}\label{sec7}
Compact astrophysical systems such as NS and BHs constitute highly sensitive and complementary laboratories for probing feebly coupled dark-sector physics. Focusing on ultralight DPs that are kinetically mixed with the visible photon but do not comprise the DM, we study resonant photon-DP oscillations in monotonic and non-monotonic plasma environments relevant to M87* and the Crab pulsar-wind Nebula. In media with a monotonic plasma mass profile, photon-DP conversion is well described by a single avoided level crossing, and the standard LZ approximation reliably captures the transition probability using only local information at the resonance. In contrast, non-monotonic plasma profiles generically lead to multiple level crossings, for which the conversion probability cannot be obtained by treating each crossing independently. In this case, coherent interference between amplitudes generated at different crossings plays a crucial role, and the resulting phase depends on the integrated eigenvalue splitting between the crossings rather than solely on local properties at the resonance. Consequently, accurate predictions in realistic astrophysical environments require a global treatment of the propagation through non-monotonic media, beyond the standard single-crossing LZ framework.

We show that physically motivated non-monotonic plasma density profiles, including log-normal profiles applicable to the M87 environment and GJ charge densities with potential QED effects in the Crab Nebula, can lead to a substantial enhancement of resonant photon-DP conversion. This enhancement translates into constraints on the kinetic-mixing parameter that are several orders of magnitude stronger than those derived assuming monotonic plasma density profiles. This enhancement arises because non-monotonic plasma profiles around M87 and Crab Nebula admit multiple resonant crossings with reduced local density gradients, increasing the adiabaticity of the transition and allowing coherent interference between successive LZ crossings, thereby substantially amplifying the photon-DP conversion probability.

Using M87 spectral data extending to the LOFAR band, we obtain $\epsilon \sim 7\times10^{-6}$ at $m_{A'} \sim 5\times10^{-7}~\mathrm{eV}$, among the strongest astrophysical limits at this scale. From the Crab SED, we further obtain an even stronger constraint, $\epsilon \simeq 8\times10^{-7}$ at $m_{A'} \simeq 4\times10^{-9}~\mathrm{eV}$ for oscillation baselines of order $10^{3}$ km, surpassing existing astrophysical limits in realistic structured plasma backgrounds. While current laboratory and cosmological bounds (e.g., Cavendish-type experiments and COBE/FIRAS limits) remain stronger by one to two orders of magnitude at comparable masses, we emphasize that plasma structuring around extreme objects offers exceptional discovery potential, with clear prospects to outperform laboratory reach in systems with larger surface magnetic fields. In particular, magnetars with surface magnetic fields as large as $B_0/B_c\sim 1000$
yield projected sensitivities that approach, and in some regions surpass, the strongest current laboratory bounds.

Non-monotonicity may also arise in angular directions, where resonance points can coalesce and the standard LZ treatment breaks down in analogy to the radial case, modifying oscillation probabilities and potentially leaving distinctive spectral signatures. The same formalism can be extended to additional photon frequency bands of M87*, as well as to high-field central engines such as magnetars powering soft gamma repeaters and gamma-ray bursts, where even stronger kinetic-mixing limits are expected from spectral analyses. Similar resonant oscillation effects can also be investigated in other plasma environments, including the intergalactic medium and the solar atmosphere, for realistic non-monotonic profiles.

Beyond photon-DP conversion, structured plasma backgrounds can also affect neutrino flavor oscillations, altering propagation phases and transition probabilities in a plasma parameter-dependent manner. Extensions to angular plasma resonances, multi-band spectral evolution, and plasma-induced modifications of both photon and neutrino oscillations constitute well-motivated future directions. We defer a dedicated analysis of these aspects to forthcoming work.

\section*{Acknowledgments}
T.K.P would like to thank Indian Association for the Cultivation of Science (IACS) for their kind hospitality, where this project was initiated. This article is based upon work from the COST Actions ``COSMIC
WISPers" (CA21106) and ``BridgeQG" (CA23130), both supported by
COST (European Cooperation in Science and Technology). 
PS acknowledges support from the University Grants Commission, Government of India, through a Senior Research Fellowship.

\bibliographystyle{JHEP}
\bibliography{refs}
\end{document}